\documentclass[11pt]{article}
\usepackage{amsmath,amssymb,color,epsfig,cite}
\usepackage{graphicx}
\usepackage{subfigure}
\usepackage{setspace}
\usepackage{physics}
\usepackage{hyperref}

\usepackage{amsfonts}

\makeatletter
\@addtoreset{equation}{section}
\makeatother

\newcommand{\be}{\begin{equation}}
\newcommand{\ee}{\end{equation}}
\newcommand{\bea}{\setlength\arraycolsep{2pt} \begin{eqnarray}}
\newcommand{\eea}{\end{eqnarray}}
\newcommand{\nn}{\nonumber}
\DeclareMathOperator{\sgn}{sgn}
\let\a=\alpha \let\b=\beta \let\g=\gamma \let\d=\delta 
    \let\k=\kappa
 \let\m=\mu \let\n=\nu \let\x=\xi \let\p=\pi 
 \let\t=\tau  \let\f=\phi  
         
\let\L=\Lambda
  \let\S=\Sigma  \let\F=\Phi \let\Y=\Psi
 \let\W=\Omega 

\def\ft#1#2{{\textstyle{\frac{\scriptstyle #1}{\scriptstyle #2} } }}
\def\fft#1#2{{\frac{#1}{#2}}}

\def\0{{\sst{0}}}
\def\1{{\sst{1}}}
\def\2{{\sst{2}}}
\def\3{{\sst{3}}}
\def\4{{\sst{(4)}}}
\def\5{{\sst{(5)}}}
\def\6{{\sst{(6)}}}
\def\7{{\sst{(7)}}}
\def\8{{\sst{(8)}}}
\def\sst#1{{\scriptscriptstyle #1}}

\def\im{{{\rm i\,}}}

\numberwithin{equation}{section}

\begin{document}

\begin{center}
{\large {\bf Kappa vacua: \\Infinite number of new vacua in two-dimensional quantum field theory}}

\vspace{15pt}
\vspace{15pt}
{\large  Arash Azizi}

\vspace{15pt}

{\it The Institute for Quantum Science and Engineering,\\
Texas A\&M University, College Station, TX 77843, USA}

Email: sazizi@tamu.edu

\vspace{30pt}

\underline{ABSTRACT}
\end{center}

\vspace{25pt}

We uncover an infinite number of vacua in two-dimensional quantum field theory, the Klein-Gordon field for simplicity, by conceiving a new mode that is classified by a real positive parameter $\k$. We show each mode has a distinct vacuum, say $\k$-vacuum. This new mode is a generalization of the  Unruh-Minkowski mode. Moreover, the Minkowski and Rindler vacua are special cases of the $\k$-vacuum for $\k=1$ and $\k \rightarrow \infty$, respectively.

\pagebreak

\section{Introduction}

The longtime goal of theoretical physics is to find a consistent theory of quantum gravity, which ultimately describes our physical world. Although we are far from achieving the goal, startling phenomena like Hawking radiation \cite{Hawking75} and the Unruh effect \cite{Unruh76} already provide some insights for us. These effects not only invoke fundamental questions like  the Hawking information paradox \cite{Hawking76}  and its possible resolutions \cite{Penington19,Almheiri2019,
west_coast,east_coast}, but have also generated interest in disparate areas, e.g., quantum optics \cite{Scully2003,Scully2022}.

In addition to the above-mentioned theoretical endeavors, their  experimental realizations have been pursued in analogue systems, since both effects are so small, which makes them inaccessible to current technologies. For instance, a particle should have a technologically impossible uniform acceleration of $10^{20} \fft {\text{m}}{{\text{s}^2}}$ in Minkowski vacuum to feel an Unruh temperature of 1~K. Hence, an alternative approach to verifying these effects experimentally can be an analogue gravity \cite{Barcelo:2005fc,Gooding:2020scc,langen2013local,torres2017rotational}.

The cornerstone of these phenomena is the existence of several vacua, rather than the unique vacuum of the more usual quantum field theory in Minkowski spacetime \cite{Fulling73}. The standard description of quantum field theory prescribes writing a quantum field, say spin zero Klein-Gordon field, as a linear combination of the solutions of the equations of motion, i.e., field modes, associated with annihilation and creation operators. The crucial point of which mode should be associated with the former or the latter operators is determined by the Klein-Gordon inner product. The positive (negative) norm mode is associated with annihilation (creation) operators.

Writing down a field in terms of distinct modes yields Bogoliubov transformations, relating annihilation operators of a specific mode to {\it both} annihilation and creation operators of another, and hence, the vacuum of the former mode is distinct from the vacuum of the latter. Consequently, it is crucial to find which mode is associated to annihilation and creation operators. 

Unruh in his pioneering work \cite{Unruh76} introduced a novel  mode whose form is similar to the Rindler one, but whose vacuum state is Minkowski instead of Rindler. It turns out this mode is very useful to study an Unruh-DeWitt detector \cite{Unruh76,Einstein100} (or simply a two-level atom) with a uniform acceleration in a flat spacetime. Unruh and Wald \cite{UnruhWald84} showed  that the detector, starting from the ground state in the Minkowski vacuum, emits an Unruh-Minkowski mode. Counter intuitively, the mode is residing more in the opposite Rindler wedge, causally disconnected from  the wedge where the detector moves. One may realize this bizarre result in a setup of two uniformly accelerated atoms, one with a positive and one with a negative acceleration, and confirm the Unruh and Wald result \cite{Svidzinsky21prl,Svidzinsky21prr}.

Here, we introduce a generalization of the Unruh-Minkowski mode, containing a positive real parameter $\k$, where  Rindler and Unruh-Minkowski modes can be found by $\k \rightarrow \infty$ and $\k=1$ respectively. The fascinating aspect of this new mode is it yields infinitely many distinct vacua, which one may call $\k$-vacua. Obviously, Rindler and Minkowski vacua are special cases of the $\k$-vacuum for $\k \rightarrow \infty$ and $\k=1$ respectively.

We consider a massless Klein-Gordon field in $1+1$ dimensions, with the metric of $ds^2=-dt^2+dx^2$. Here, we set $c=1$, and consider the $(-,+)$ convention for the metric. It is more convenient to use the Minkowski light-cone coordinates $u=t-x$ and $v=t+x$. The field equation is  $\Box \F=0$, or in terms of light-cone coordinates  $\ft{\partial}{\partial u}\ft{\partial}{\partial v}\F=0$, and can be solved simply by $\F(u,v)=\F(u)+\Y(v)$, where $\F(u)$ and $\Y(v)$ are general functions indicating right- and left-moving waves respectively. Consider the change of coordinates
\be
u=-\ft1a e^{-a(\t-\x)}\,, \qquad \qquad 
v=\ft1a e^{a(\t+\x)}\,.
\ee
One may call $(\t,\x)$ Rindler coordinates \cite{Rindler66} and the metric shall be written as $ds^2= e^{2 a\, \x}\,(-d \t^2 +d \x^2)$.   For positive (negative) $a$, we have $u<0$ ($u>0$) and $v>0$ ($v<0$)  and the associated region in spacetime diagram is called Rindler right (left) wedge.

\section{Combining the opposite sign norm Rindler modes}

The Klein-Gordon inner product \cite{DEWITT1975, Wald:1975kc} is exploited here to distinguish the positive- and negative-norm modes associated to the annihilation and creation operators respectively. The inner product between two modes $\F_1$ and $\F_2$ reads
\begin{equation}
\left\langle \f_1, \f_2\right\rangle=
-\im \int_{\S}\, \sqrt{-g}\,
d\S^\m \left(\f_1^{*} \, \partial_\m \f_2
-\partial_\m  \f_1^{*}\, \f_2\right)\,, \label{KGinnerproduct}
\end{equation}
where $\S$ is the appropriate Cauchy hypersurface. Note, we have used the convention of $(-,+, \cdots, +)$ for the Minkowski metric. Moreover, in the $1+1$ dimensional Minkowski spacetime, if one chooses the light-cone coordinates $(u,v)$, then the inner product (\ref{KGinnerproduct}) becomes  
\begin{equation}
\left\langle f, g\right\rangle=\im \int_{-\infty}^{\infty} d v\left(f^{*} \frac{\partial}{\partial v} g-\frac{\partial}{\partial v} f^{*} g\right) \,, \quad
\left\langle f, g\right\rangle=\im \int_{-\infty}^{\infty} d u\left(f^{*} \frac{\partial}{\partial u} g-\frac{\partial}{\partial u} f^{*} g\right)\,,\label{innerprod(u,v)}
\end{equation}
where the first (second) term above is written for constant $u$ ($v$) surfaces.

From the Klein-Gordon inner product, one may derive the positive-norm Rindler modes, for a right-moving wave, in the left and right wedges, as $\F(u,\W)=\theta(u)\,\fft1{\sqrt{4\p\W}}\,u^{-\im \W}$ and $\F(u,\W)=\theta(-u)\,\fft1{\sqrt{4\p\W}}\,(-u)^{\im \W}$ respectively. Here $\W$ is a positive real number. The similar expressions can be found for the left-moving wave in terms of $v$.

The positive-norm, right-moving wave Unruh-Minkowski mode may be written as 
\be
\F_{UM}(u,\W)=
\theta(u)\,\fft{e^{-\ft{\p\W}2}\,
u^{\im \W}}{\sqrt{8\p\W \sinh{(\p\W)}}} 
+ \theta(-u)\,\fft{e^{\ft{\p\W}2}\,
(-u)^{\im \W}}{\sqrt{8\p\W \sinh{(\p\W)}}}\,,
\ee
where $\W$ is any real number. As we have emphasized, the mode's vacuum is the Minkowski one;  however, its form resembles the Rindler modes. This fascinating feature is possible thanks to the combination of Rindler modes in the right and left wedges. One then wonders about the existence of other modes constructed by a  similar method, i.e., combination of Rindler modes in the right and left wedges, but, with the most general coefficients. The coefficients shall be fixed by demanding the orthonormality of the modes, with respect to the Klein-Gordon inner product. It turns out a family of such modes exists. The surprising feature of the new modes is that their vacua are distinct from the Minkowski. There are two different ways to find the new modes: Combination of the same and the opposite sign norm Rindler modes. 

An ansatz for the opposite sign norm is
\begin{equation}
\F(u,\W)=\a(\Omega)\theta(-u)(-u)^{\im \Omega}
+\b(\Omega)\theta(u) u^{\im \Omega}\,, \label{ansatz2}
\end{equation}
where we have considered here the right-moving wave. The inner product of two modes reads
\be
\langle \F(u,\W), \F(u,\W')\rangle= 4\p\W\, \Big( \abs{\a(\W)}^2-\abs{\b(\W)}^2 \Big) \,\d(\W-\W')
\,. \label{ip5}
\ee
Therefore, the orthonormality of the norm indicates
\be
4\p\W\, \Big( \abs{\a(\W)}^2-\abs{\b(\W)}^2 \Big)=1\,,\label{constraint2}
\ee
where $\W$ can be either positive or negative real parameter.

Furthermore, the inner product of the mode and its conjugate becomes
\begin{align}
\langle \F(u,\W),& \F^*(u,\W')\rangle= 4\p\W \,\d(\W+\W')\, \Big( \a^*(\Omega)\a^*(\Omega')-\b^*(\Omega)\b^*(\Omega') \Big)\,.\label{ip7}
\end{align}
This should be zero and it imposes a constraint on $\a(\W)$ and $\b(\W)$. Namely,
\be
\a(\Omega)\a(-\Omega)-\b(\Omega)\b(-\Omega)=0\,,\label{constraint3}
\ee
for all $\W$. To solve the constraints, first start with (\ref{constraint2}). Without loss of generality, the following  ansatz can be introduced:
\be
\a(\W)=\frac{e^{\fft{\k\p \Omega}2+\im \theta}}{\sqrt{8 \pi \Omega\, \sinh{(\k \p\Omega)}}} \,,
\qquad \qquad
\b(\W)=\frac{e^{-\fft{\k\p \Omega}2+\im \f}}{\sqrt{8 \pi \Omega\, \sinh{(\k \p\Omega)}}}\,,
\ee
where $\k$ is an arbitrary positive real number, and $\theta$ and $\f$ are any real numbers. Since $\k>0$,  $8 \pi \Omega\, \sinh{(\k \p\Omega)}$ is always positive. Next, imposing the second constraint (\ref{constraint3}) implies $\theta=\f$. 
Hence, $e^{\im \theta}$ is an overall phase that can be ignored. Therefore, we have
\be
\a(\W)=\frac{e^{\fft{\k\p \Omega}2}}{\sqrt{8 \pi \Omega\, \sinh{(\k \p\Omega)}}} \,,
\qquad \qquad
\b(\W)=\frac{e^{-\fft{\k\p \Omega}2}}{\sqrt{8 \pi \Omega\, \sinh{(\k \p\Omega)}}}\,. \label{ab}
\ee

Having found $\a$ and $\b$, now the new mode can be written as
\be
\F(u,\W,\k)= \fft1{\sqrt{8 \pi \Omega\, \sinh{(\k\p \Omega)}}}
\left\{\theta(-u)(-u)^{\im \Omega}\, 
e^{\frac{\k \pi \Omega}{2}}
+\theta(u) u^{\im \Omega}\, e^{-\frac{\k \pi \Omega}{2}} \right\} \,. \label{modeopp}
\ee

In addition to this, the field of the right-moving wave can be written as 
\be
\F(u)=\int_{-\infty}^{\infty} d\W \Big(\F(u,\W,\k) {\cal A}_{\W,\k} + \F^*(u,\W,\k) {\cal A}^{\dagger}_{\W,\k} \Big)\,,\label{fieldopp}
\ee
where we have used ${\cal A}_{\W,\k}$ to denote the annihilation operator for the $\k$-mode in the opposite sign norm scenario. 

One of the interesting features of the new mode introduced here is that the Rindler and Unruh-Minkowski modes can be found as special cases. Namely, set $\k\rightarrow\infty$ in (\ref{modeopp}) and the mode becomes $\F(u,\k\rightarrow\infty)= \fft1{\sqrt{4 \pi \W}}
\theta(-u)\, (-u)^{ \im \W}$ for positive $\W$, and for negative $\W$, after setting $\W \rightarrow -\W$, the mode yields   $\F(u,\k\rightarrow\infty)= \fft1{\sqrt{4 \pi \W}}
\theta(u)\, u^{ -\im \W}$. These are the familiar Rindler modes in the right and left wedges respectively. Furthermore, taking $\k=1$ in (\ref{modeopp}) easily gives the Unruh-Minkowski mode
\be
\F(u,\W,\k=1)= \fft1{\sqrt{8 \pi \Omega\, \sinh{(\p \Omega)}}}
\left\{\theta(-u)(-u)^{\im \Omega}\, e^{\frac{\pi \Omega}{2}}
+\theta(u) u^{\im \Omega}\, e^{-\frac{\pi \Omega}{2}}\right\} \,. \label{UM}
\ee

The next step is to find the Bogoliubov transformation between two distinct modes. Specifically, one may write $\F(u)$ in terms of ${\cal A}_{\W,\k}$ and ${\cal A}^{\dagger}_{\W,\k}$ as written in (\ref{fieldopp}), and also equally, in terms of ${\cal A}_{\W',\k'}$ and ${\cal A}^{\dagger}_{\W',\k'}$. After comparing $\theta(-u) (-u)^{\im \L}$ and  $\theta(u) u^{\im \L}$ in both expressions, the Bogoliubov transformation then becomes
\be
\begin{pmatrix}
{\cal A}_{\L,\k'}\\
{\cal A}^{\dagger}_{-\L,\k'}
\end{pmatrix} = 
\fft{\sgn{(\L)}}{\sqrt{{\sinh{(\k\p \L)}\sinh{(\k'\p \L)}}}}
\begin{pmatrix}
\sinh{\big(\ft{(\k+\k')\pi \L}2 \big)}\qquad
&\sinh{\big(\ft{(\k'-\k)\pi \L}2\big)}\\
\sinh{\big(\ft{(\k'-\k)\pi \L}2\big)}\qquad
&\sinh{\big(\ft{(\k+\k')\pi \L}2 \big)}
\end{pmatrix}\,
\begin{pmatrix}
{\cal A}_{\L,\k}\\
{\cal A}^{\dagger}_{-\L,\k}
\end{pmatrix}
\,. \label{bogolfinal}
\ee
This crucial relation has very important consequences. First, let's find the Bogoliubov transformation between Rindler and Unruh-Minkowski. Setting $\k\rightarrow \infty$ and $\k'=1$ yields 
\bea
{A}_{\L} &=&\fft1{\sqrt{1- e^{-2\p \L}}}\,
\Big(b_{R\L}- e^{-\p\L}\,b^{\dagger}_{L\L}\Big)\,,\nn\\
{A}_{-\L}&=&\fft1{\sqrt{1- e^{-2\p \L}}}\,
\Big(b_{L\L}- e^{-\p\L}\,b^{\dagger}_{R\L}\Big)\,,
\label{bogolUM-R}
\eea
where $\L>0$. This is exactly as expressed in \cite{UnruhWald84}. Here the Unruh-Minkowski annihilation operators  are indicated by ${\cal A}_{\W,1} ={ A}_{\W}$ for all real $\W$, while the Rindler annihilation operators are represented by ${\cal A}_{\W,\infty}= b_{R\W}$ and ${\cal A}_{-\W,\infty}=b_{L\W}$ for the right and left wedges respectively, and $\W>0$.

Second, (\ref{bogolfinal}) vividly shows that vacua for modes associated with different $\k$ are distinct. The vacuum annihilated by ${\cal A}_{\L,\k}$ can be labeled by $\ket{0_{\k}}$, and may be called the $\k$-vacuum. The off-diagonal term in (\ref{bogolfinal}) expresses that $\ket{0_{\k}}$ and $\ket{0_{\k'}}$ are distinct.

One may also find the Bogoliubov transformation between a $\k$-mode and a Minkowski plane wave. This can be achieved by   setting the field for the right-moving wave written in terms of a $\k$-mode  (\ref{modeopp}, \ref{fieldopp}), and the one written in Minkowski plane wave form, i.e.,  $\F(u)=\int_{0}^{\infty}
 \fft{d\n}{\sqrt{4 \p \n}}\, 
 \big(a_{\n}\,e^{-\im \n \, u}
 + a^\dagger_{\n}\,e^{\im \n \, u} \big)$, equal to each other to find the following relation:

\begin{align}
a_{\n} = \sqrt{\ft{\n}{\p}}  \int_{-\infty}^{\infty} d\W 
\fft{\im}{\sqrt{2 \pi \Omega\, \sinh{(\k\p \Omega)}}} 
\Bigg\{&-\n^{-(1+\im \Omega)}\,\Gamma(1+\im \Omega)
\sinh{\Big(\frac{(\k+1) \pi \Omega}{2}\Big)}
\,{\cal A}_{\W,\k} \nn\\
&- \n^{-(1-\im \Omega)}\,\Gamma(1-\im \Omega)
\sinh{\Big(\frac{(\k-1) \pi \Omega}{2}\Big)}
\,{\cal A}^{\dagger}_{\W,\k} \Bigg\} 
 \,. \label{bogolkappa-planewave}
\end{align}
The above relation clearly shows that for the case of $\k=1$, the creation operator of the $\k$-mode does not contribute, hence, its vacuum is Minkowski. However, this is not a surprising result, since we have shown that the $\k=1$ case is actually the Unruh-Minkowski mode, where the vacuum state is Minkowski.

\section{Combining the same sign norm Rindler modes}

In the similar fashion as the previous case, an ansatz for the same sign norm, and again, right-moving wave is as follows
\begin{equation}
\F(u,\W)=\a(\Omega)\theta(-u)(-u)^{\im \Omega}
+\b(\Omega)\theta(u) u^{-\im \Omega}\,. \label{ansatz1}
\end{equation}
Note for positive $\W$ both terms above have positive norm, while for negative $\W$, both of them have negative norm. 

The mode orthonormality, i.e., $\left\langle \F(u,\W), \F(u,\W')\right\rangle= \d(\W-\W')$, yields 

\be
4\p\W\, \Big( \abs{\a(\W)}^2+\abs{\b(\W)}^2 \Big) =1\,, \label{samenormconst}
\ee
and hence, in contrast to the previous case, $\W$ can be just a  positive real parameter. Furthermore, $\langle \F(u,\W), \F^*(u,\W')\rangle   
=\Big(\a^*(\Omega)\a^*(\Omega')
+\b^*(\Omega)\b^*(\Omega')\Big) 4\p \W\,\d(\W+\W')$ which is automatically zero, since $\W$ and $\W'$ are both positive. The constraint (\ref{samenormconst}) may be solved as 
\be
\a(\W)=\frac{e^{\fft{\k\p \Omega}2+\fft{\im \g}2}}{\sqrt{8 \pi \Omega\, \cosh{(\k \p\Omega)}}} \,,
\qquad \qquad
\b(\W)=\frac{e^{-\fft{\k\p \Omega}2-
\fft{\im \g}2}}{\sqrt{8 \pi \Omega\, \cosh{(\k \p\Omega)}}}\,,
\ee
where $\k$ and $\g$ are any real numbers. The above form is chosen up to an overall phase. Therefore, one may write the following expression for the modes for the same positivity norm: 
\be
\F(u,\W,\k,\g)= \frac{1}{\sqrt{8 \pi \Omega\, \cosh{(\k \p\Omega)}}}\,
\left\{\theta(-u)(-u)^{\im \Omega}\, 
e^{\fft{\k\p \Omega}2+\fft{\im \g}2}
+\theta(u) u^{-\im \Omega}\, 
e^{\fft{-\k\p \Omega}2-\fft{\im \g}2}\right\} \,. \label{modesame}
\ee

The field can be written as 
\be
\F(u)=\int_{0}^{\infty} d\W \Big(\F(u,\W,\k,\g) {\cal A}_{\W,\k,\g} + \F^*(u,\W,\k,\g) {\cal A}^{\dagger}_{\W,\k,\g} \Big)\,, \label{fieldsame}
\ee
where we have used ${\cal A}_{\W,\k,\g}$ to denote the annihilation operator for  the same positivity norm $(\k,\g)$-mode.

The next step is to find the Bogoliubov transformation between  $(\k,\g)$ and $(\k',\g')$-modes. It then implies
\bea
\fft1{\sqrt{8 \pi \L\, \cosh{(\k\p \L)}}} 
\,e^{\frac{\k \pi \L}{2}}\, e^{\frac{\im \gamma}{2}}\,{\cal A}_{\L,\k,\g} &=&
\fft1{\sqrt{8 \pi \L\, \cosh{(\k'\p \L)}}} 
\,e^{\frac{\k' \pi \L}{2}}\, e^{\frac{\im \gamma'}{2}}\,{\cal A}_{\L,\k',\g'}\,, \nn\\
\fft1{\sqrt{8 \pi \L\, \cosh{(\k\p \L)}}} 
\,e^{-\frac{\k \pi \L}{2}}\, e^{-\frac{\im \gamma}{2}}\,{\cal A}_{\L,\k,\g} &=&
\fft1{\sqrt{8 \pi \L\, \cosh{(\k'\p \L)}}} 
\,e^{-\frac{\k' \pi \L}{2}}\, e^{-\frac{\im \gamma'}{2}}\,{\cal A}_{\L,\k',\g'}\,. \label{bogol2same}
\eea
The above expressions simply yield $\k=\k'$ and $\g=\g'$.  Consequently if there exists a mode in the same sign norm, then $\k$ and $\g$ would be unique.

Now, let us find the Bogoliubov transformation between the same positivity modes and the opposite positivity modes of the previous section. Using (\ref{modeopp}), (\ref{fieldopp}), (\ref{modesame}), and (\ref{fieldsame}), and comparing two fields in the same and the opposite positivity modes yields  
\be
\fft{e^{\frac{-\k \pi \L}{2}
+\frac{\im \gamma}{2}}}{\sqrt{8 \pi \L\, \cosh{(\k\p \L)}}} {\cal A}^{\dagger}_{\L,\k,\g}
=\fft1{\sqrt{8 \pi \L\, \sinh{(\k'\p \L)}}} 
\Big( e^{-\frac{\k' \pi \L}{2}}\,  {\cal A}_{\L,\k'} 
+e^{\frac{\k' \pi \L}{2}}\,  {\cal A}^{\dagger}_{-\L,\k'} 
\Big)\,, \label{oppsame1}
\ee
for reading off $\theta(u)u^{\im \L}$ in two fields, and also 
\be
\fft{e^{\frac{\k \pi \L}{2}
+\frac{\im \gamma}{2}}}{\sqrt{8 \pi \L\, \cosh{(\k\p \L)}}} {\cal A}_{\L,\k,\g}
=\fft1{\sqrt{8 \pi \L\, \sinh{(\k'\p \L)}}} 
\Big( e^{\frac{\k' \pi \L}{2}}\,  {\cal A}_{\L,\k'} 
+e^{\frac{-\k' \pi \L}{2}}\,  {\cal A}^{\dagger}_{-\L,\k'} 
\Big)\,,\label{oppsame2}
\ee
for reading off  $\theta(-u)(-u)^{\im \L}$. In the above expressions we have considered $\L>0$.

Now it is clear while the left-hand sides of (\ref{oppsame2}) and Hermitian conjugate of (\ref{oppsame1}) are proportional, the right-hand sides are not. Therefore, one concludes the same sign norm mode cannot exist even for a unique value of $\k$ and $\g$.

As a result,  the same sign norm mode, expressed in terms of $(\k,\g)$-mode, does not exist.

\section{Conclusion}

A genuine mode which is classified by a continuous positive real parameter $\k$, and hence is named $\k$-mode, is introduced. Moreover, Bogoliubov transformation is used to demonstrate the vacua of these modes are indeed distinct. Henceforth, our work uncovers an infinite set of new vacua in two-dimensional quantum field theory. The $\k$-mode is found by a combination of the Rindler modes in the right and left wedges with unknown coefficients. These coefficients can be fixed by demanding the Klein-Gordon inner product orthonormality. Two distinct cases of the opposite and the same sign norm have been investigated, and it turns out, modes exist only in the former case. Furthermore, two already well-known modes, i.e., Rindler and Unruh-Minkowski, are special cases of $\k$-mode for $\k\rightarrow\infty$ and $\k=1$ respectively. Finally, finding a physical interpretation for the $\k$-vacuum is an ongoing project.

\section*{Acknowledgments}

AA is grateful to Girish Agarwal, Jonathan Ben-Benjamin, Yusef Maleki, Anatoly Svidzinsky, and especially Marlan Scully and Bill Unruh  for useful and illuminating discussions. AA thanks Reed Nessler for helpful comments on the manuscript.

This work was supported by the Air Force Office of Scientific
Research (Grant No. FA9550-20-1-0366 DEF), the
Robert A. Welch Foundation (Grant No. A-1261), and the National
Science Foundation (Grant No. PHY-2013771).

\bibliographystyle{jhep}
\bibliography{Kappa}
\end{document}